\documentclass[english,aps,preprint]{revtex4}
\usepackage[T1]{fontenc}
\usepackage[usenames]{color}
\usepackage{amsmath}
\usepackage{cancel}
\usepackage{graphicx}
\usepackage{babel}
\usepackage{ulem}
\usepackage{multirow}

\allowdisplaybreaks

\begin{document}

\title{From QCD sum rules to HQET sum rules: Heavy-quark limit of four-quark operator matrix elements}

\author{
En-Qi Wu$^{1}$,
Yi-Peng Xing$^{2}$,
Chen-Hang Weng$^{1}$,
Zhen-Xing Zhao$^{1,3}$~\footnote{Email: zhaozx19@imu.edu.cn}
}

\affiliation{
$^{1}$ School of Physical Science and Technology, Inner Mongolia University, Hohhot 010021, China\\
$^{2}$ School of Physics, Henan Normal University, Xinxiang 453007, Henan, People's Republic of China\\
$^{3}$ Research Center for Quantum Physics and Technologies, Inner Mongolia University, Hohhot 010021, China
}

\begin{abstract}
Heavy quark expansion theory provides the standard theoretical framework for understanding the lifetimes of weakly decaying heavy-flavor hadrons. Within this framework, the matrix elements of four-quark operators play a crucial role in explaining the lifetime differences among hadrons containing the same heavy quark.
In this work, we calculate these matrix elements at leading order using full QCD sum rules, with particular emphasis on deriving the corresponding HQET sum rules by taking the heavy-quark limit. While this limit is straightforward for most contributions, it turns out to be rather nontrivial for certain ones. Numerical analyses show that the results obtained from the two approaches are consistent with each other. We further clarify the origin of the large discrepancies reported in the literature between full QCD sum rule and HQET sum rule results.
This work contributes to a deeper understanding of the relationship between the two sum-rule approaches and is expected to facilitate future applications of QCD and HQET sum rules in the study of heavy-flavor hadrons. The findings presented here may be of interest to both practitioners of QCD sum rules and researchers working on effective field theories.
\end{abstract}

\maketitle

\section{Introduction}

In 2018, the LHCb Collaboration updated the measurement of the lifetime
of the charmed baryon $\Omega_{c}^{0}$ \cite{LHCb:2018nfa}
\begin{equation}
\tau(\Omega_{c}^{0})=268\pm24\pm10\pm2\,\text{fs},
\end{equation}
which is nearly four times larger than the world average value in PDG2018 \cite{Tanabashi:2018oca}
\begin{equation}
\tau(\Omega_{c}^{0})=69\pm12\,\text{fs}.
\end{equation}
In 2021, LHCb confirmed its 2018 measurement \cite{LHCb:2021vll}, and in 2022, the Belle-II Collaboration reported a similar result \cite{Belle-II:2022plj}.

To date, the standard theoretical framework for studying the lifetimes
of weakly decaying heavy-flavor hadrons is the Heavy Quark Expansion (HQE). 
For a detailed theoretical exposition, see Subsec.~\ref{subsec:HQE} below. On the
theoretical side, several attempts have been made to resolve the $\Omega_{c}$
lifetime puzzle; some of these can be found in Refs. \cite{Cheng:2018rkz,Gratrex:2022xpm,Cheng:2023jpz}.
In these works, the calculations of the four-quark operator matrix
elements (FOMEs) are all based on various quark models, and a lattice QCD calculation is still lacking. In fact, FOMEs can also be computed using QCD sum rules (QCDSR).

QCDSR is a non-perturbative method that connects hadron properties
to QCD vacuum condensates via dispersion relations. In Ref.~\cite{Zhao:2021lzd},
we calculated the FOMEs of $\Lambda_{b}$ using full QCD sum rules (full-QCDSR). However, when comparing our results with those of Heavy Quark Effective Theory sum rules (HQET-SR) in Ref.~\cite{Colangelo:1996ta},
we found that our results are several times larger. For example, Ref.~\cite{Zhao:2021lzd} gives
\begin{equation}
L_{1}=-0.143\pm0.028,
\end{equation}
while Ref.~\cite{Colangelo:1996ta} gives
\begin{equation}
L_{1}=-0.033\pm0.017.
\end{equation}
The definition of $L_{1}$ is provided in Eq.~(\ref{eq:Li}) below. This discrepancy partially arises from the different renormalization scales adopted in the two calculations, but that is clearly not the dominant origin. One of the primary purposes of the present work is to identify the key causes underlying this inconsistency.

In fact, taking the heavy-quark limit of the FOMEs calculated from full-QCDSR in Ref.~\cite{Zhao:2021lzd} can reproduce the results of HQET-SR given in Ref.~\cite{Colangelo:1996ta}. 
This fact can be understood as follows. The FOMEs in full QCD and HQET can be calculated within their respective theories using the operator product expansion (OPE) technique. The results can be formally written as
\begin{align}
\langle\Lambda_{b}|O_{6}^{{\rm QCD}}|\Lambda_{b}\rangle 
&= \sum_{i} C_{i}^{{\rm QCD}}\langle O_{i}\rangle,\nonumber \\
\langle\Lambda_{b}|O_{6}^{{\rm HQET}}|\Lambda_{b}\rangle 
&= \sum_{i} C_{i}^{{\rm HQET}}\langle O_{i}\rangle.
\end{align}
Here, $C_{i}^{{\rm QCD}}$ and $C_{i}^{{\rm HQET}}$ denote the corresponding Wilson coefficients, while $O_i$ generally do not involve heavy-quark fields. Since in the present work we consider only the leading order in $\alpha_s$ and the leading power in $1/m_Q$, the matching from $O_{6}^{{\rm QCD}}$ to $O_{6}^{{\rm HQET}}$ is trivial. Therefore, the Wilson coefficients $C_{i}^{{\rm HQET}}$ should be exactly the leading power in $1/m_Q$ of $C_{i}^{{\rm QCD}}$.
Another major motivation of this work is to demonstrate this point explicitly, thereby indirectly validating our approach for evaluating the spectral density of baryonic three-point correlation functions using the cutting rule. Evaluating three-point correlation functions within QCD sum rules is nontrivial, especially when we note that that some treatments in the existing literature are incorrect. In our previous works \cite{Shi:2019hbf,Zhao:2020mod,Zhao:2021sje,Xing:2021enr}, we computed three-point correlation functions based on full-QCDSR and extracted the form factors for several weak decay processes. The method adopted here and in Ref.~\cite{Zhao:2021lzd} for calculating FOMEs is essentially a natural generalization of the framework established in those studies.

In addition to our work, several recent studies have also employed the cutting rule to compute the spectral density of three-point correlation functions; see, e.g., Refs.~\cite{Zhang:2023nxl,Zhang:2024ick,Zhang:2024asb,Lu:2025gol,Yu:2026tbk}. It is worth pointing out that if radiative corrections are taken into account, one may still need to resort to integration by parts (IBP) methods for multi-loop calculations. Regarding the calculation of baryon three-point correlation functions, no relevant studies have been reported so far. For the calculation of meson three-point correlation functions, the most recent literature can be found, for example, in Ref.~\cite{Black:2024bus}.

Several additional remarks are worth emphasizing. First, the discussions in this work were originally inspired by Refs. \cite{Shuryak:1981fza,Colangelo:1995qp}, which drew our attention to the fact that taking the heavy-quark limit of full-QCDSR results can yield HQET-SR results. Second, as a method close to first principles, QCD sum rules characterize hadronic observables using only universal vacuum condensate parameters. From a practical perspective, the dominant systematic uncertainty originates from the continuum threshold parameter $s_{0}$ introduced in quark-hadron duality. At least in principle, the optimal value of $s_{0}$ can be determined by requiring the observable to be insensitive to the Borel parameter. Nevertheless, it is often difficult for conventional methods to accurately estimate such systematic uncertainties. Recently developed inverse problem methods may offer a promising pathway to improve this situation \cite{Li:2020ejs,Xiong:2022uwj,Zhao:2024drr}.

The remainder of this article is organized as follows. Sec.~II briefly describes the theoretical framework of the HQE, followed by the formulations of two-point and three-point correlation functions as well as the procedure for taking the heavy-quark limit. A subsection has also been added to discuss the determination of the parameters in QCDSR analyses. Sec.~III presents the numerical results and the error estimates, and Sec.~IV provides a summary and conclusions.

\section{Theoretical Framework}

\subsection{Heavy quark expansion}
\label{subsec:HQE}

The Heavy Quark Expansion (HQE) \cite{Khoze:1983yp,Bigi:1991ir,Bigi:1992su,Blok:1992hw,Blok:1992he,Neubert:1997gu,Uraltsev:1998bk,Bigi:1995jr} provides a systematic framework for describing inclusive weak decays of heavy-flavor hadrons. It is based on the operator product expansion (OPE) in inverse powers of the heavy-quark mass $m_{Q}$, which allows for a separation of short-distance and long-distance effects. For a relatively recent and systematic review, the reader is referred to Ref.~\cite{Lenz:2014jha}.

Using the optical theorem, the total decay width of a hadron $H_{Q}$ containing a heavy quark $Q$ can be expressed as 
\begin{equation}
\Gamma(H_{Q})=\frac{\langle H_{Q}|2\,{\rm Im}{\cal T}|H_{Q}\rangle}{2\,M_{H}},
\end{equation}
where the transition operator ${\cal T}$ is defined by
\begin{equation}
{\cal T}=i\int d^{4}x\,T[{\cal L}_{W}(x){\cal L}_{W}^{\dagger}(0)],
\end{equation}
with ${\cal L}_{W}$ being the weak effective Lagrangian that governs the heavy quark decay. The transition operator can then be evaluated using the OPE technique:
\begin{equation}
2\,{\rm Im}{\cal T}=\frac{G_{F}^{2}m_{Q}^{5}}{192\pi^{3}}\xi\left(c_{3}\bar{Q}Q+\frac{c_{5}}{m_{Q}^{2}}\bar{Q}g_{s}\sigma GQ+\frac{c_{6}}{m_{Q}^{3}}(\bar{Q}\Gamma q\,\bar{q}\Gamma^{\prime}Q)+\cdots\right),
\end{equation}
where $\xi$ denotes the relevant CKM matrix elements. The dimension-3 and dimension-5 terms respectively describe the quasi-free heavy quark decay and the corresponding corrections from the gluon field, while the dimension-6 four-quark operators are the primary source of the lifetime differences among hadrons containing the same heavy quark $Q$.

The relevant baryon matrix elements can be parameterized in a model-independent way \cite{Cheng:2018rkz}. For the $\Lambda_{b}$ baryon, the spectator effects encoded in the four-quark operators can be parameterized as 
\begin{align}
\langle\Lambda_{b}\lvert(\bar{b}q)_{V-A}(\bar{q}b)_{V-A}\lvert\Lambda_{b}\rangle= & f_{B_{q}}^{2}m_{B_{q}}m_{\Lambda_{b}}L_{1},\nonumber \\
\langle\Lambda_{b}\lvert(\bar{b}q)_{S-P}(\bar{q}b)_{S+P}\lvert\Lambda_{b}\rangle= & f_{B_{q}}^{2}m_{B_{q}}m_{\Lambda_{b}}L_{2},\nonumber \\
\langle\Lambda_{b}\lvert(\bar{b}^{\alpha}q^{\beta})_{V-A}(\bar{q}^{\beta}b^{\alpha})_{V-A}\lvert\Lambda_{b}\rangle= & f_{B_{q}}^{2}m_{B_{q}}m_{\Lambda_{b}}L_{3},\nonumber \\
\langle\Lambda_{b}\lvert(\bar{b}^{\alpha}q^{\beta})_{S-P}(\bar{q}^{\beta}b^{\alpha})_{S+P}\lvert\Lambda_{b}\rangle= & f_{B_{q}}^{2}m_{B_{q}}m_{\Lambda_{b}}L_{4},\label{eq:Li}
\end{align}
where $(\bar{q}_{1}q_{2})_{V-A}\equiv\bar{q}_{1}\gamma_{\mu}(1-\gamma_{5})q_{2}$ and $(\bar{q}_{1}q_{2})_{S\pm P}\equiv\bar{q}_{1}(1\pm\gamma_{5})q_{2}$. The $L_{i}$ are dimensionless nonperturbative parameters that encode the long-distance QCD dynamics. In phenomenological analyses, relations among these parameters, such as $L_{3}=-\tilde{B}L_{1}$, are often introduced.

In the following, we will demonstrate how these baryon matrix elements can be calculated within the framework of full-QCDSR, and how the corresponding HQET-SR results can be derived by taking the heavy-quark limit. 
First, we consider the two-point correlation function. This serves two purposes: it is simpler, and the pole residue extracted from two-point correlation function is an indispensable input for the subsequent calculations. Then, we will discuss how to extract the parameters $L_{1,2}$ using three-point correlation functions, as well as how to take the heavy quark limit.

\subsection{Two-point correlation function}

The QCD sum rule analysis of the two-point correlation function for $\Lambda_b$ has been extensively studied in the literature; see, for example, Ref.~\cite{Wang:2010fq}.
For the sake of completeness, we nevertheless briefly outline the main steps below.

The construction of sum rules begins with the definition of interpolating currents for the hadrons involved. For $\Lambda_{b}$, we adopt the following interpolating current:
\begin{equation}
J=\epsilon_{ijk}(u_{i}^{T}C\gamma_{5}d_{j})b_{k},
\end{equation}
where $i,j,k$ are color indices, and $C$ is the charge conjugation matrix. The pole residue characterizes the coupling of the interpolating current to the baryon state. To extract the pole residue, we consider the standard two-point correlation function
\begin{equation}
\Pi(p)=i\int d^{4}x\,e^{ip\cdot x}\langle0\lvert T\{J(x)\bar{J}(0)\}\lvert0\rangle.\label{eq:2pt}
\end{equation}

On the hadronic side, after inserting a complete set of hadronic states, the correlation function can be written as
\begin{equation}
\Pi^{\mathrm{had}}(p)=\lambda_{+}^{2}\frac{\cancel{p}+M_{+}}{M_{+}^{2}-p^{2}}+\lambda_{-}^{2}\frac{\cancel{p}-M_{-}}{M_{-}^{2}-p^{2}}+\cdots,
\end{equation}
where we have also included the contribution from the negative-parity baryon. Here $M_{\pm}$ and $\lambda_{\pm}$ denote the masses and pole residues of the positive- and negative-parity baryons, respectively. The pole residues $\lambda_{\pm}$ are defined by
\begin{align}
\langle0|J|\Lambda_{b+}(p,s)\rangle & =\lambda_{+}u(p,s),\nonumber \\
\langle0|J|\Lambda_{b-}(p,s)\rangle & =(i\gamma_{5})\lambda_{-}u(p,s).
\end{align}

\begin{figure}
\centering
\includegraphics[width=0.5\linewidth]{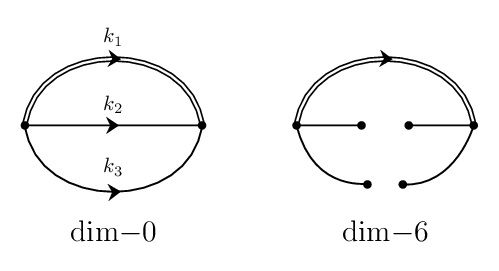}
\caption{Feynman diagrams for the perturbative and four-quark condensate contributions to the two-point correlation function considered in this work.}
\label{fig:feynman-2pt}
\end{figure}

On the QCD side, we evaluate the correlation function in Eq.~(\ref{eq:2pt}) using the OPE. 
Contributions from operators up to dimension-6 are included. The quark condensate and mixed condensate contributions vanish, while the gluon condensate contribution is numerically highly suppressed and therefore neglected~\cite{Wang:2010fq}. The dominant contributions included in our calculation are illustrated in  Fig.~\ref{fig:feynman-2pt}. 
The results can be formally expressed as
\begin{equation}
\Pi^{\mathrm{QCD}}(p)=A_{1}(p^{2})\cancel{p}+A_{2}(p^{2}),
\end{equation}
where the coefficients $A_{i}$ can be further written in the form of dispersion relations:
\begin{equation}
A_{i}(p^{2})=\int ds\frac{\rho_{i}(s)}{s-p^{2}}.
\label{eq:dispersion_2pt}
\end{equation}
In Eq.~(\ref{eq:dispersion_2pt}), $\rho_{i}$ represents the spectral density, which can be evaluated using, for example, the Cutkosky cutting rules.
The explicit expressions for these spectral densities can be found in Appendix~\ref{sec:rho-2pt}.

Using the quark-hadron duality ansatz and performing the Borel transform, one obtains the following sum rule:
\begin{equation}
(M_{+}+M_{-})\lambda_{+}^{2}\exp(-M_{+}^{2}/T_{+}^{2})=\int^{s_{0}}ds(M_{-}\rho_{1}+\rho_{2})\exp(-s/T_{+}^{2}),\label{eq:SR_2pt}
\end{equation}
where $s_{0}$ and $T_{+}^{2}$ are the continuum threshold parameter and the Borel parameter, respectively. For more details, see, e.g., Ref.~\cite{Zhao:2021lzd}. 

From Eq.~(\ref{eq:SR_2pt}), one can derive the mass formula for $\Lambda_{b}$:
\begin{equation}
M_{+}^{2}=\frac{\int^{s_{0}}ds(M_{-}\rho_{1}+\rho_{2})\,s\,\exp(-s/T_{+}^{2})}{\int^{s_{0}}ds(M_{-}\rho_{1}+\rho_{2})\exp(-s/T_{+}^{2})}.\label{eq:mass}
\end{equation}

The sum rule analysis is performed on Eq.~(\ref{eq:SR_2pt}). Once the optimal values of $s_{0}$ and $T_{+}^{2}$ are determined, they are substituted into Eq.~(\ref{eq:mass}) to compute the baryon mass. In this work, the extracted baryon mass serves as one of the criteria for validating the sum rule.

\subsubsection*{The heavy-quark limit}

In this subsubsection, we derive the heavy-quark limit of the full-QCDSR for the pole residue. Our goal is to establish a direct connection between full-QCDSR and HQET-SR. 

Before performing the Borel transform, Eq.~(\ref{eq:SR_2pt}) reverts to
\begin{equation}
(M_{+}+M_{-})\lambda_{+}^{2}\frac{1}{M_{+}^{2}-p^{2}}=\int^{s_{0}}ds\,(M_{-}\rho_{1}+\rho_{2})\frac{1}{s-p^{2}}.\label{eq:SR_2pt_pre}
\end{equation}
We introduce the following variable replacements:
\begin{equation}
M_{\pm}=(m_{b}+\Delta_{\pm}),\quad p^{2}=(m_{b}+\omega)^{2},\quad s=(m_{b}+\sigma)^{2},\quad s_{0}=(m_{b}+\omega_{c})^{2},
\end{equation}
where $\Delta_{+(-)}$ is the binding energy for $\Lambda_{b}(1/2^{+(-)})$, and we adopt the notation conventions of Ref.~\cite{Colangelo:1996ta}. To leading power in $1/m_{b}$, Eq.~(\ref{eq:SR_2pt_pre}) can be rewritten as
\begin{equation}
\lambda_{+}^{2}\frac{1}{(\Delta_{+}-\omega)}=\int_{0}^{\omega_{c}}d\sigma\,\rho_{\mathrm{HQL}}(\sigma)\,\frac{1}{(\sigma-\omega)},\label{eq:SR_2pt_HQL_pre}
\end{equation}
where $\rho_{\mathrm{HQL}}(\sigma)$ denotes the leading power in $1/m_{b}$ of $(M_{-}\rho_{1}+\rho_{2})$. Then, after performing the Borel transform on Eq.~(\ref{eq:SR_2pt_HQL_pre}), we obtain the sum rule in the heavy-quark limit:
\begin{equation}
\lambda_{+}^{2}\exp\Bigl(-\frac{\Delta_{+}}{E_{+}}\Bigr)=\int_{0}^{\omega_{c}}d\sigma\,\rho_{\mathrm{HQL}}(\sigma)\,\exp\Bigl(-\frac{\sigma}{E_{+}}\Bigr),\label{eq:SR_2pt_HQL}
\end{equation}
where $E_{+}$ is the new Borel parameter.

The expression for $\rho_{\mathrm{HQL}}(\sigma)$ can be obtained straightforwardly as
\begin{equation}
\rho_{\mathrm{HQL}}(\sigma)=\frac{\sigma^{5}}{20\pi^{4}}+\frac{\langle\bar{q}q\rangle^{2}}{6}\delta(\sigma),\label{eq:rho_HQL}
\end{equation}
where the first term corresponds to the perturbative contribution, and the second term arises from the four-quark condensate. In Ref.~\cite{Colangelo:1996ta}, only the perturbative term is present; our result here coincides with theirs.

One can then obtain from Eq.~(\ref{eq:SR_2pt_HQL}) the binding energy for $\Lambda_{b}$:
\begin{equation}
\Delta_{+}=\frac{\int_{0}^{\omega_{c}}d\sigma\,\rho_{\mathrm{HQL}}(\sigma)\,\sigma\,\exp(-\sigma/E_{+})}{\int_{0}^{\omega_{c}}d\sigma\,\rho_{\mathrm{HQL}}(\sigma)\,\exp(-\sigma/E_{+})}.\label{eq:binding}
\end{equation}

\subsection{Three-point correlation function}

To extract the FOMEs, we consider the following three-point correlation function:
\begin{equation}
\Pi(p_{1},p_{2})=i^{2}\int d^{4}xd^{4}ye^{-ip_{1}\cdot x+ip_{2}\cdot y}\langle0\lvert T\{J(y)\Gamma_{6}(0)\bar{J}(x)\}\lvert0\rangle,\label{eq:correlator_3pt}
\end{equation}
where $\Gamma_{6}$ denotes a four-quark operator as given in Eq.~(\ref{eq:Li}). Note that in Dirac space, $\Pi(p_{1},p_{2})$ is a $4\times4$ matrix.

Following the standard QCD sum rule procedure, the correlation function is evaluated both at the hadronic level and at the QCD level. At the hadronic level, after inserting a complete set of baryonic states and including the contribution from the negative-parity baryon, we obtain
\begin{align}
\Pi^{\mathrm{had}}(p_{1},p_{2}) & = \lambda_{+}\lambda_{+}\frac{(\cancel{p}_{2}+M_{+})(a^{++}+b^{++}\gamma_{5})(\cancel{p}_{1}+M_{+})}{(p_{2}^{2}-M_{+}^{2})(p_{1}^{2}-M_{+}^{2})}\nonumber \\
 & +\lambda_{+}\lambda_{-}\frac{(\cancel{p}_{2}+M_{+})(a^{+-}+b^{+-}\gamma_{5})(\cancel{p}_{1}-M_{-})}{(p_{2}^{2}-M_{+}^{2})(p_{1}^{2}-M_{-}^{2})}\nonumber \\
 & +\lambda_{-}\lambda_{+}\frac{(\cancel{p}_{2}-M_{-})(a^{-+}+b^{-+}\gamma_{5})(\cancel{p}_{1}+M_{+})}{(p_{2}^{2}-M_{-}^{2})(p_{1}^{2}-M_{+}^{2})}\nonumber \\
 & +\lambda_{-}\lambda_{-}\frac{(\cancel{p}_{2}-M_{-})(a^{--}+b^{--}\gamma_{5})(\cancel{p}_{1}-M_{-})}{(p_{2}^{2}-M_{-}^{2})(p_{1}^{2}-M_{-}^{2})}\nonumber \\
 & +\cdots.\label{eq:hadronic_3pt}
\end{align}
Here, $M_{+(-)}$ and $\lambda_{+(-)}$ denote the mass and pole residue of $\Lambda_{b}(1/2^{+(-)})$, respectively; the ellipsis represents the contributions from excited states and the continuum; and $a^{\pm\pm}$, $b^{\pm\pm}$ are defined through the following conventions:
\begin{align}
\langle\Lambda_{b+}(q^{\prime},s^{\prime})\lvert\Gamma_{6}\lvert\Lambda_{b+}(q,s)\rangle= & \bar{u}_{+}(q^{\prime},s^{\prime})(a^{++}+b^{++}\gamma_{5})u_{+}(q,s),\nonumber \\
\langle\Lambda_{b+}(q^{\prime},s^{\prime})\lvert\Gamma_{6}\lvert\Lambda_{b-}(q,s)\rangle= & \bar{u}_{+}(q^{\prime},s^{\prime})(a^{+-}+b^{+-}\gamma_{5})(i\gamma_{5})u_{-}(q,s),\nonumber \\
\langle\Lambda_{b-}(q^{\prime},s^{\prime})\lvert\Gamma_{6}\lvert\Lambda_{b+}(q,s)\rangle= & \bar{u}_{-}(q^{\prime},s^{\prime})(i\gamma_{5})(a^{-+}+b^{-+}\gamma_{5})u_{+}(q,s),\nonumber \\
\langle\Lambda_{b-}(q^{\prime},s^{\prime})\lvert\Gamma_{6}\lvert\Lambda_{b-}(q,s)\rangle= & \bar{u}_{-}(q^{\prime},s^{\prime})(i\gamma_{5})(a^{--}+b^{--}\gamma_{5})(i\gamma_{5})u_{-}(q,s).\label{eq:ab}
\end{align}
In Eq.~(\ref{eq:ab}), the factors of $(i\gamma_{5})$ are introduced for convenience. For the forward scattering matrix elements in Eq.~(\ref{eq:Li}), one can show that
\begin{equation}
\langle\Lambda_{b+}(q,s)\lvert\Gamma_{6}\lvert\Lambda_{b+}(q,s)\rangle=2\,a^{++}\,m_{\Lambda_{b}},\label{eq:12}
\end{equation}
which implies that only the parameter $a^{++}$ in Eq.~(\ref{eq:ab}) is relevant to the forward scattering matrix element. For further details, see Ref.~\cite{Zhao:2021lzd}.

\begin{figure}
\centering
\includegraphics[width=1.0\linewidth]{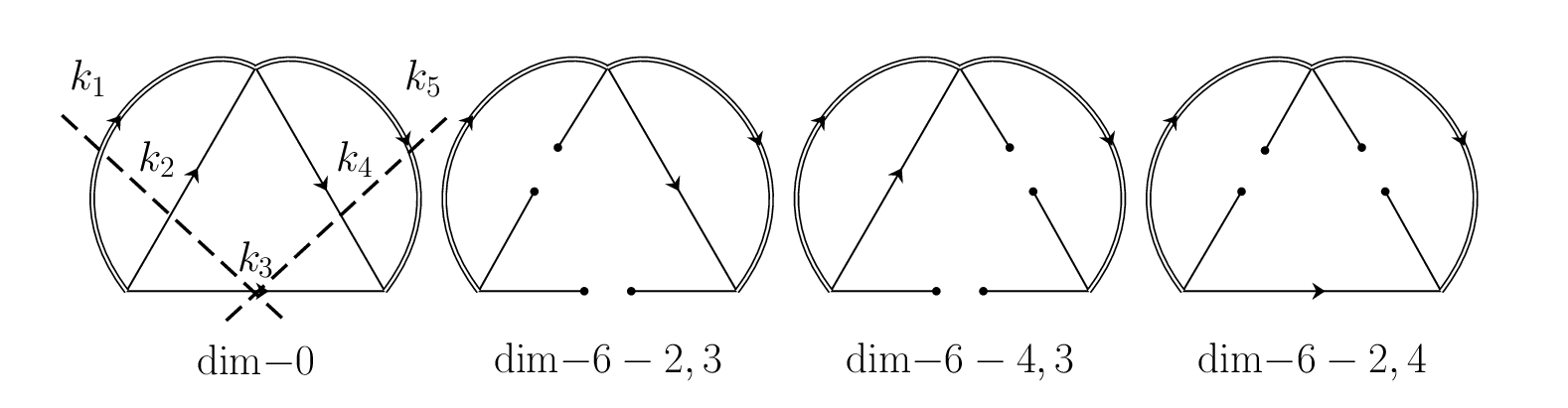}
\caption{Feynman diagrams for the perturbative and four-quark condensate contributions to the three-point correlation functions considered in this work.}
\label{fig:feynman-3pt}
\end{figure}

At the QCD level, the OPE calculation is performed, where contributions from operators up to dimension-6 are included. 
Analogous to the case of the two-point correlation function, 
the quark condensate and mixed condensate contributions vanish, while the gluon condensate contribution is highly suppressed~\cite{Zhao:2021lzd}. The dominant contributions included in our calculation are illustrated in Fig.~\ref{fig:feynman-3pt}. 
The calculation results of the correlation function in Eq.~(\ref{eq:correlator_3pt}) can be formally expressed as
\begin{align}
\Pi^{\mathrm{QCD}}(p_{1},p_{2})= & \{A_{1},A_{2},A_{3},A_{4},A_{5},A_{6},A_{7},A_{8}\}\nonumber \\
 & .\{\cancel{p}_{2}\cancel{p}_{1},\,\cancel{p}_{2},\,\cancel{p}_{1},\,1,\,\cancel{p}_{2}\gamma_{5}\cancel{p}_{1},\,\cancel{p}_{2}\gamma_{5},\,\gamma_{5}\cancel{p}_{1},\,\gamma_{5}\},\label{eq:QCD_3pt}
\end{align}
where the coefficients $A_{i}$ are further written in terms of double dispersion relations:
\begin{equation}
A_{i}(p_{1}^{2},p_{2}^{2},q^{2})=\int^{\infty}ds_{1}\int^{\infty}ds_{2}\frac{\rho^{A_{i}}(s_{1},s_{2},q^{2})}{(s_{1}-p_{1}^{2})(s_{2}-p_{2}^{2})},\label{eq:22}
\end{equation}
and the spectral density functions $\rho^{A_{i}}(s_{1},s_{2},q^{2})$ are evaluated using the Cutkosky cutting rules. 
The corresponding cutting rules is illustrated in the perturbative diagram of Fig.~\ref{fig:feynman-3pt}. 
The explicit expressions for these spectral densities can be found in Appendix~\ref{sec:rho-3pt}.

The sum rules are constructed by equating Eq.~(\ref{eq:hadronic_3pt}) and Eq.~(\ref{eq:QCD_3pt}) and then invoking quark-hadron duality to eliminate the contributions from excited states and the continuum. Equating the coefficients of identical Dirac structures yields eight equations for the eight unknown parameters $a^{\pm\pm}$ and $b^{\pm\pm}$. In particular, after performing the Borel transform, one obtains
\begin{equation}
a^{++}=\frac{\{M_{-}^{2},M_{-},M_{-},1\}\cdot\{\mathcal{B}A_{1},\mathcal{B}A_{2},\mathcal{B}A_{3},\mathcal{B}A_{4}\}}{\lambda_{+}^{2}(M_{+}+M_{-})^{2}}\exp\left(\frac{2M_{+}^{2}}{T^{2}}\right),\label{eq:SR_3pt}
\end{equation}
where $\mathcal{B}A_{i}$ are the doubly Borel-transformed coefficients:
\begin{equation}
\mathcal{B}A_{i}=\int^{s_{0}}ds_{1}\int^{s_{0}}ds_{2}\,\rho^{A_{i}}(s_{1},s_{2},q^{2})\exp\left(-\frac{s_{1}+s_{2}}{T^{2}}\right),\label{eq:BAi}
\end{equation}
with $s_{0}$ and $T^{2}$ being the continuum threshold parameter and the Borel parameter, respectively.

\subsubsection*{Heavy-quark limit}

We now derive the heavy-quark limit of the full-QCDSR for the FOMEs. It should be emphasized that this procedure is nontrivial for three-point correlation functions, where complex mathematics may be involved. 

Before performing the Borel transform, Eq.~(\ref{eq:SR_3pt}) reverts to the following form:
\begin{align}
 & \lambda_{+}^{2}(M_{+}+M_{-})^{2}a^{++}\frac{1}{(M_{+}^{2}-p_{1}^{2})(M_{+}^{2}-p_{2}^{2})}\nonumber \\
= & \int^{s_{0}}ds_{1}\int^{s_{0}}ds_{2}\,\{M_{-}^{2},M_{-},M_{-},1\}.\{\rho^{A_{1}},\rho^{A_{2}},\rho^{A_{3}},\rho^{A_{4}}\}\frac{1}{(s_{1}-p_{1}^{2})(s_{2}-p_{2}^{2})}.\label{eq:SR_3pt_pre}
\end{align}
We introduce the following variable replacements:
\begin{equation}
M_{\pm}=(m_{b}+\Delta_{\pm}),\quad p_{1(2)}^{2}=(m_{b}+\omega_{1(2)})^{2},\quad s_{1(2)}=(m_{b}+\sigma_{1(2)})^{2},\quad s_{0}=(m_{b}+\omega_{c})^{2}.
\end{equation}
Then, to leading power in $1/m_{b}$, Eq.~(\ref{eq:SR_3pt_pre}) can be rewritten as
\begin{equation}
\lambda_{+}^{2}a^{++}\frac{1}{(\Delta_{+}-\omega_{1})(\Delta_{+}-\omega_{2})}=\int^{\omega_{c}}d\sigma_{1}\int^{\omega_{c}}d\sigma_{2}\,\rho_{\mathrm{HQL}}(\sigma_{1},\sigma_{2})\frac{1}{(\sigma_{1}-\omega_{1})(\sigma_{2}-\omega_{2})},\label{eq:SR_3pt_HQL_pre}
\end{equation}
where $\rho_{\mathrm{HQL}}(\sigma_{1},\sigma_{2})$ denotes the leading power in $1/m_{b}$ of $\{M_{-}^{2},M_{-},M_{-},1\}.\{\rho^{A_{1}},\rho^{A_{2}},\rho^{A_{3}},\rho^{A_{4}}\}$. After performing the double Borel transform on Eq.~(\ref{eq:SR_3pt_HQL_pre}), we obtain
\begin{equation}
\lambda_{+}^{2}a^{++}\exp\left(-\frac{2\Delta_{+}}{F}\right)=\int^{\omega_{c}}d\sigma_{1}\int^{\omega_{c}}d\sigma_{2}\,\rho_{\mathrm{HQL}}(\sigma_{1},\sigma_{2})\exp\left(-\frac{\sigma_{1}+\sigma_{2}}{F}\right),\label{eq:SR_3pt_HQL}
\end{equation}
where $F$ is the new Borel parameter. In Ref.~\cite{Colangelo:1996ta}, $F=2E_{+}$ is adopted. In this work, however, we treat these two Borel parameters independently.

We now present the heavy quark limit expressions for $L_{1}$. The spectral density function $\rho_{\mathrm{HQL}}(\sigma_{1},\sigma_{2})$ is obtained as
\begin{equation}
\rho_{\mathrm{HQL}}(\sigma_{1},\sigma_{2})=\rho^{(\mathrm{pert})}(\sigma_{1},\sigma_{2})+\rho^{(D=6)}(\sigma_{1},\sigma_{2})\,\langle\bar{q}q\rangle^{2}+\cdots,
\end{equation}
where the perturbative term and the four-quark condensate term are respectively given by 
\begin{align}
\rho^{(\mathrm{pert})}(\sigma_{1},\sigma_{2}) & =-\frac{3}{8\pi^{6}}\Bigl\{\theta(\sigma_{1}-\sigma_{2})\sigma_{2}^{5}\Bigl(\frac{\sigma_{2}^{2}}{105}-\frac{\sigma_{1}\sigma_{2}}{30}+\frac{\sigma_{1}^{2}}{30}\Bigr)+(\sigma_{1}\leftrightarrow\sigma_{2})\Bigr\},\nonumber \\
\rho^{(D=6)}(\sigma_{1},\sigma_{2}) & =-\frac{1}{24\pi^{2}}\Bigl\{\sigma_{2}^{2}\delta(\sigma_{1})+\sigma_{1}^{2}\delta(\sigma_{2})+\sigma_{1}\sigma_{2}\delta(\sigma_{1}-\sigma_{2})\Bigr\}.
\end{align}
The first two terms of $\rho^{(D=6)}$ are associated with the four-quark condensate diagrams dim-6-2,3 and dim-6-4,3, while the last term corresponds to dim-6-2,4, as can be seen in Fig.~\ref{fig:feynman-3pt}. Most of the calculations of taking the heavy quark limit are straightforward, while that for dim-6-2,4 is rather nontrivial, involving taking some limit of a theta function to yield a delta function, as can be seen in Appendix~\ref{sec:HQL-624}.
Our analytical expressions obtained here are consistent with those in Ref.~\cite{Colangelo:1996ta}, noting that the operator definition in the present work differs from that in Ref.~\cite{Colangelo:1996ta} by a factor of $(-4)$.

\subsection{Some notes on the parameter selection in QCD sum rules}
\label{Subsec:QCDSR_para}

The continuum threshold parameter and the Borel parameter are two important 
parameters in QCD sum rules. In this work, we attempt to treat all these quantities 
merely as ``parameters'' as much as possible, without introducing additional 
artificial assumptions. Moreover, different sum rules are treated completely 
independently. This means that the optimal continuum threshold parameters and 
Borel parameters determined from different sum rules are, in principle, 
completely independent of each other. Therefore, even for the determination of 
the same physical quantity, the optimal continuum threshold parameters obtained 
from full-QCDSR and HQET-SR may deviate slightly from the ideal relation
\begin{equation}
\omega_{c}=\sqrt{s_{0}}-m_{b}.
\label{eq:s0_omega_c}
\end{equation}

We first derive some general conclusions about the Borel parameters. For the 
two-point correlation function, by comparing Eq.~(\ref{eq:SR_2pt}) with 
Eq.~(\ref{eq:SR_2pt_HQL}), we obtain the exact correspondence between the Borel 
parameters in full-QCDSR and HQET-SR:
\begin{equation}
T_{+}^{2}=2m_{b}E_{+}.
\label{eq:T2_E}
\end{equation}
Similarly, for the three-point correlation function, by comparing 
Eq.~(\ref{eq:SR_3pt}) with Eq.~(\ref{eq:SR_3pt_HQL}), we obtain
\begin{equation}
T^{2}=2m_{b}F.
\label{eq:T2_F}
\end{equation}
We expect that both $E_{+}$ and $F$ are of 
${\cal O}(\Lambda_{\rm QCD})$. Therefore, we conclude that the typical Borel 
parameter in the full theory is of order
${\cal O}(2m_{b}\Lambda_{\rm QCD})$.

We further provide some rough numerical estimates. For the two-point HQET-SR, 
Eq.~(\ref{eq:SR_2pt_HQL}) contains the exponential factor
\begin{equation}
\exp\left(-\frac{\sigma-\Delta_{+}}{E_{+}}\right).
\end{equation}
Taking
\begin{equation}
E_{+}=\Delta_{+}=0.9\,{\rm GeV},
\end{equation}
we immediately obtain, from Eq.~(\ref{eq:T2_E}), the corresponding Borel 
parameter in full-QCDSR:
\begin{equation}
T_{+}^{2}\approx10\,{\rm GeV}^{2}.
\end{equation}
For the three-point HQET-SR, Eq.~(\ref{eq:SR_3pt_HQL}) contains the exponential 
factor
\begin{equation}
\exp\left(
-\frac{(\sigma_{1}-\Delta_{+})+(\sigma_{2}-\Delta_{+})}{F}
\right).
\end{equation}
This motivates the estimate
\begin{equation}
F=2\Delta_{+}=2\times0.9\,{\rm GeV}.
\end{equation}
Using Eq.~(\ref{eq:T2_F}), we immediately obtain the corresponding Borel 
parameter in full-QCDSR:
\begin{equation}
T^{2}\approx20\,{\rm GeV}^{2}.
\end{equation}

Consequently, based on the above reasonable estimates, we obtain
\begin{align}
F_{\rm 3pt} &=2E_{+,{\rm 2pt}},
\label{eq:F_2E}\\
T_{\rm 3pt}^{2} &=2T_{+,{\rm 2pt}}^{2}.
\label{eq:T2_Tp2}
\end{align}
The relation in Eq.~(\ref{eq:F_2E}) is exactly the prescription adopted in Ref.~\cite{Colangelo:1996ta}. In addition, the relation in 
Eq.~(\ref{eq:T2_Tp2}) has also been used in some previous works.

The above estimation of the scale of the Borel parameters plays the role of 
the Borel window in the QCDSR analysis of this work. Subsequently, we determine 
the optimal continuum threshold parameter by requiring that the physical 
quantity exhibits maximal stability around the Borel window.

\section{Numerical results}

\subsection{Inputs}

For the numerical analysis, we adopt the following input parameters. The up and down quark masses are negligible and are therefore set to zero. The bottom quark mass is taken as the pole mass $m_{b}=4.80\pm0.06$ GeV \cite{Penin:1998kx,Colangelo:2000dp}. The masses of the $\Lambda_{b}(1/2^{\pm})$ baryons are chosen as $M_{+}=m_{\Lambda_{b}(1/2^{+})}=5.620$ GeV and $M_{-}=m_{\Lambda_{b}(1/2^{-})}=5.912$ GeV, respectively \cite{ParticleDataGroup:2024cfk}. For the quark condensate, we adopt the standard value $\langle\bar{q}q\rangle\,(1\,{\rm GeV})=(-0.24\pm0.01\,{\rm GeV})^{3}$ \cite{Colangelo:2000dp}.

\subsection{The pole residue, mass and binding energy}

\begin{figure}[htbp]
\centering
\includegraphics[width=1.0\linewidth]{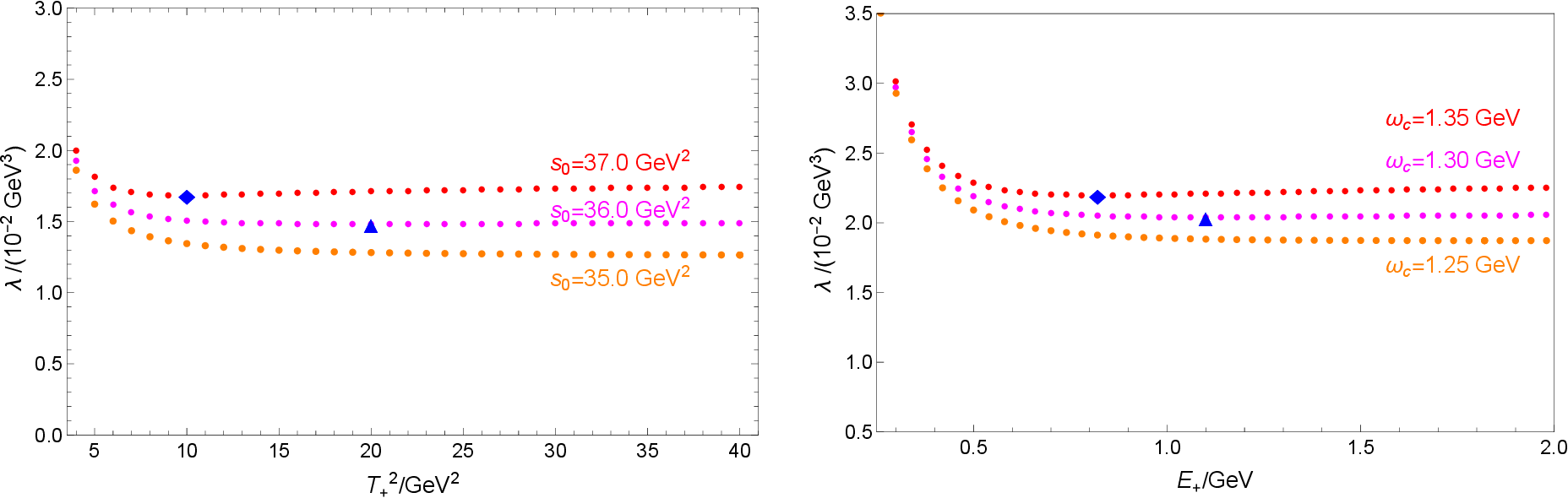}
\caption{The pole residue in full-QCDSR (left) and HQET-SR (right).}
\label{fig:pole_residue}
\end{figure}

\begin{figure}[htbp]
\centering
\includegraphics[width=1.0\linewidth]{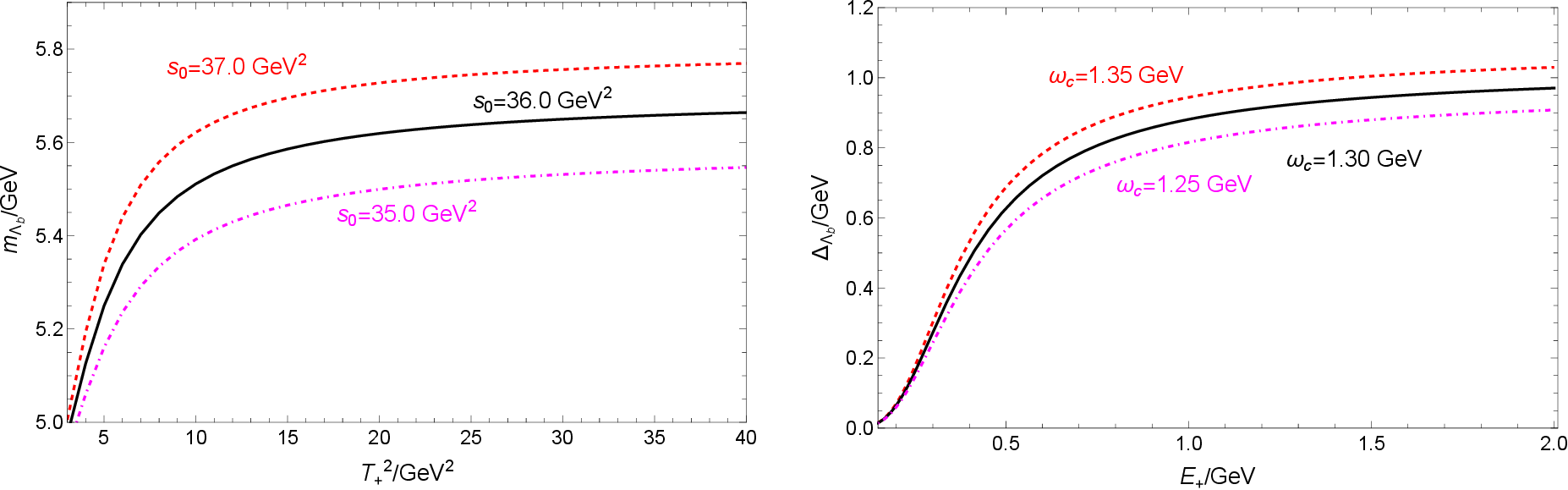}
\caption{The mass in full-QCDSR (left) and the binding energy in HQET-SR (right). Up to the next-to-leading power in $1/m_{b}$, we have $m_{\Lambda_{b}}=m_{b}+\Delta_{\Lambda_{b}}$.}
\label{fig:binding}
\end{figure}

Within the full-QCDSR framework, we study the pole residue and mass of $\Lambda_{b}$, while in HQET-SR, we investigate the pole residue and binding energy. The dependence of these quantities on the Borel parameter is shown in Figs.~\ref{fig:pole_residue} and \ref{fig:binding}, and the corresponding numerical
results are listed in Table \ref{Tab:results}. For a complete uncertainty analysis, please refer to the following Subsec.~\ref{subsec:uncertainty} and Table~\ref{Tab:errs} therein.


Some comments are in order. 
\begin{itemize}
\item In QCDSR analyses, the Borel window represents the reasonable range of the Borel parameter and serves as a crucial reference. 
In Subsec.~\ref{Subsec:QCDSR_para} above, we model-independently obtain the Borel window $T_{+}^{2}\sim{\cal O}(2m_{b}\Lambda_{\rm QCD})$ in full-QCDSR,
and $E_{+}\sim{\cal O}(\Lambda_{\rm QCD})$ in HQET-SR.
\item In the full-QCDSR analysis, the pole residue of the $\Lambda_b$ baryon is 
determined from Eq.~(\ref{eq:SR_2pt}). By requiring that the physical quantity 
of interest (the pole residue) is as insensitive as possible to the 
unphysical Borel parameter around the Borel window, we obtain the optimal value 
$s_{0}=36.0\,{\rm GeV}^{2}$. The uncertainty induced by the continuum 
threshold parameter is conservatively estimated by varying this optimal value 
by $1.0\,{\rm GeV}^{2}$. 
When calculating the pole residue using Eq.~(\ref{eq:SR_2pt}), the experimental 
masses $M_{\pm}$ of the positive- and negative-parity $\Lambda_b$ baryons are 
taken as inputs. Meanwhile, the theoretical prediction for the positive-parity 
baryon mass obtained from Eq.~(\ref{eq:mass}) is
$m_{\Lambda_b(1/2^{+})}^{\rm theo}\approx5.619\,{\rm GeV}$.
This result is only used as an auxiliary criterion to verify the reliability 
of the sum rule in Eq.~(\ref{eq:SR_2pt}).
\item Following the same logic, in the HQET-SR analysis, when calculating the pole 
residue using Eq.~(\ref{eq:SR_2pt_HQL}), the binding energy of the $\Lambda_b$ 
baryon,
$\Delta_{\Lambda_b}=0.9\pm0.1\,{\rm GeV}$,
is taken as an input. This value is extracted from the relation
$m_{\Lambda_b}=m_b+\Delta_{\Lambda_b}$.
We obtain the optimal effective continuum threshold parameter
$\omega_c=1.30\,{\rm GeV}$. The corresponding theoretical prediction for the 
binding energy obtained from Eq.~(\ref{eq:binding}) is
$\Delta_{\Lambda_b(1/2^{+})}^{\rm theo}\approx0.900\,{\rm GeV}$.
Similarly, this result only serves as an auxiliary criterion to test the 
reliability of the sum rule in Eq.~(\ref{eq:SR_2pt_HQL}).
\item The HQET-SR result for the pole residue is consistent with that from full-QCDSR within 
uncertainties. The residual difference can naturally be attributed to 
higher-order corrections in $1/m_b$.
\end{itemize}

\begin{table}
\caption{Numerical results for the pole residue, mass, binding energy and $L_{1,2}$. In the heavy quark limit, $L_{2} = (-1/2) L_{1}$ holds exactly in this work. For full-QCDSR (HQET-SR), we have considered the uncertainties arising from the $b$-quark mass (binding energy) and the threshold parameter $s_0$ (effective threshold parameter $\omega_c$). For a complete uncertainty analysis, please refer to the following Subsec.~\ref{subsec:uncertainty} and Table~\ref{Tab:errs} therein.}
\label{Tab:results}
\begin{tabular}{c|c|c|c|c|c}
\hline \hline
Method & $\lambda_{+}/{\rm GeV}^{3}$ & $m_{\Lambda_{b}}(\Delta_{\Lambda_{b}})/{\rm GeV}$ & $L_{1}$ & $L_{2}$ & $\tilde{B}$\tabularnewline
\hline 
Full-QCDSR & $0.0148\pm0.0026$ & $5.619\pm0.003$ & $-0.119\pm0.041$ & $0.0591\pm0.0235$ & $1$\tabularnewline
\hline 
HQET-SR & $0.0204\pm 0.0037$ & $0.900\pm 0.006$ & $-0.159\pm 0.044$ & - - & $1$\tabularnewline
\hline \hline
\end{tabular}
\end{table}

\subsection{The four-quark operator matrix elements}

\begin{figure}
\centering
\includegraphics[width=1.0\linewidth]{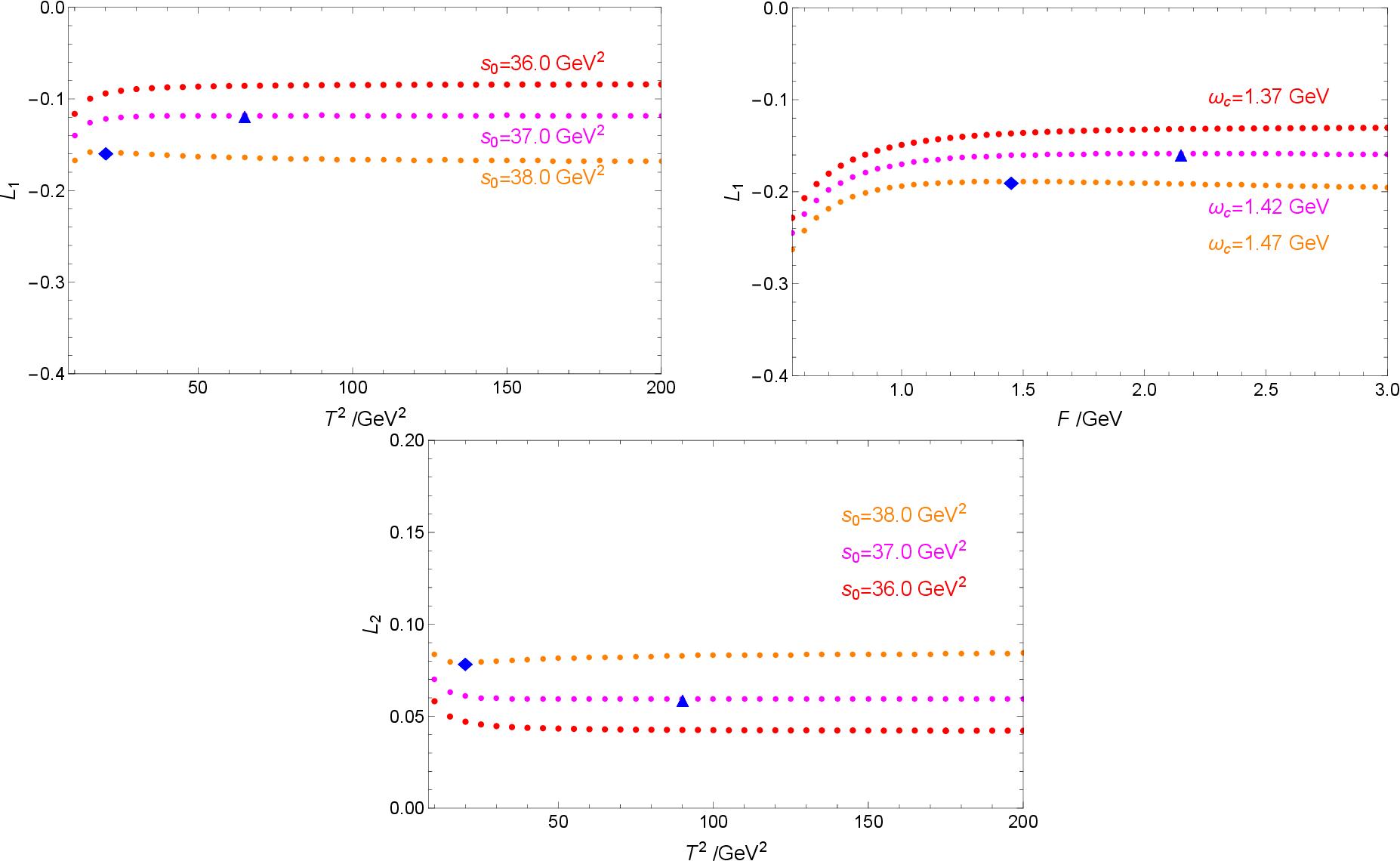}
\caption{$L_{1}$ calculated in full-QCDSR (top left) and HQET-SR (top right), and $L_{2}$ calculated in full-QCDSR (bottom). In the heavy quark limit, $L_{2} = (-1/2) L_{1}$ holds exactly in this work.}
\label{fig:Li}
\end{figure}

The FOMEs in Eq.~(\ref{eq:Li}) are parameterized in terms of a set of hadronic parameters. In the numerical analysis, we have used $m_{B}=5.280~\text{GeV}$ and $f_{B}=186~\text{MeV}$ \cite{Cheng:2018rkz}. 
Corresponding numerical results are shown in Fig.~\ref{fig:Li} and Table ~\ref{Tab:results}. 
For a complete uncertainty analysis, please refer to the following Subsec.~\ref{subsec:uncertainty} and Table~\ref{Tab:errs} therein.

Some comments are in order.
\begin{itemize}
\item The spectral density in Eq.~(\ref{eq:BAi}) also depends on $q^{2}\equiv(p_{1}-p_{2})^{2}$. For the forward scattering matrix elements of interest, $q^{2}\to 0^{-}$ is understood. 
\item We do not perform separate calculations for $L_{3}$ and $L_{4}$, because for the perturbative contributions and four-quark condensate contributions considered in this work, the first and third operators, as well as the second and fourth operators in Eq.~(\ref{eq:Li}), differ only by a factor of $(-1)$ in color space. That is,  $L_{3}=-L_{1}$ and $L_{4}=-L_{2}$ hold exactly in this work.
\item In the heavy quark limit, $L_{2} = (-1/2) L_{1}$ holds exactly for the perturbative contribution and four-quark condensate contribution considered in this work. Therefore, in Fig.~\ref{fig:Li} and Table~\ref{Tab:results}, we do not show the results for $L_{2}$ in HQET-SR. 
\item From Table~\ref{Tab:results}, we can see that the full-QCDSR and HQET-SR results for $L_{1}$ are in good agreement. The small deviations are expected to be of ${\cal O}(1/m_{b})$. 
\item In this work, we treat the threshold parameter as an independent parameter, allowing different sum rules to take different threshold values. In full-QCDSR, the continuum threshold $s_0$ for the pole residue is varied in the range $s_0 = 35.0\sim 37.0\ \text{GeV}^2$, while for $L_{1,2}$ it is varied in $s_0 = 36.0\sim 38.0\ \text{GeV}^2$. In HQET-SR, the effective threshold $\omega_c$ for the pole residue is varied in $\omega_c = 1.25\sim 1.35\ \text{GeV}$, whereas for $L_1$ it is varied in $\omega_c = 1.37\sim 1.47\ \text{GeV}$. 
The ideal continuum threshold parameters should, of course, be identical or 
satisfy the relation in Eq.~(\ref{eq:s0_omega_c}), and 
the uncertainty arising from this difference should also be of 
${\cal O}(1/m_b)$.
\end{itemize}

Upon comparison, we find that $L_{1,2}$ obtained in present work are close to those in our earlier work \cite{Zhao:2021lzd}, but differ significantly from those in Ref.~\cite{Colangelo:1996ta}. In more detail, the main reasons include the following aspects:
\begin{itemize}
\item First, Eq.~(39) of Ref.~\cite{Colangelo:1996ta} adopts the so-called local quark-hadron duality, where only the perturbative contribution is retained. In the present work, we also include the four-quark condensate contribution. In fact, for typical values of the threshold parameter and the Borel parameter in our HQET-SR analysis, the four-quark condensate contribution is only about $20\%$ of the perturbative contribution for the two-point correlation function, while for the three-point correlation function, the four-quark condensate contribution is about twice the perturbative contribution. To identify a stability region, we are forced to raise the value of the effective threshold parameter $\omega_c$. In this work, the effective threshold parameter for $L_{1}$ is varied in the range $\omega_{c}=1.37\sim1.47$ GeV, which is higher than the range $\omega_{c}=1.1\sim1.3$ GeV adopted in Ref.~\cite{Colangelo:1996ta}. As a result, the central value of $L_{1}$ is significantly increased by several times compared to that in Ref.~\cite{Colangelo:1996ta}. 
\item Second, comparing the full-QCDSR calculation in this work with our earlier work \cite{Zhao:2021lzd}, the most significant difference lies in the choice of the $b$ quark mass. In this work, we adopt the pole mass of the $b$ quark, $m_{b}^{{\rm pole}}=4.8\,{\rm GeV}$, whereas in Ref.~\cite{Zhao:2021lzd}, we adopted the $\overline{{\rm MS}}$ scheme, i.e., $m_{b}^{\overline{{\rm MS}}}=4.18\,{\rm GeV}$. The pole residue is sensitive to the heavy quark mass; whereas it seems that $L_{1,2}$ do not appear to be so.
\item Third, as also mentioned in the Introduction, another important difference between the full-QCDSR calculation in this work and our earlier work \cite{Zhao:2021lzd} is the choice of the renormalization scale. In this work, we choose the scale to be $1$ GeV; while in Ref.~\cite{Zhao:2021lzd}, our scale was $\mu=m_{b}$. 
\item Fourth, in Ref.~\cite{Colangelo:1996ta}, the authors adopted the assumption that the Borel parameter for the three-point correlation function is twice that for the two-point correlation function. In contrast, in the present work, we treat the two Borel parameters independently. Of course, this uncertainty is very small, as can be seen from Eq.~(\ref{eq:F_2E}) together with Figs.~\ref{fig:pole_residue} and \ref{fig:Li}.
\end{itemize}

\subsection{The uncertainty analysis}
\label{subsec:uncertainty}

In the following, we consider all the major sources of uncertainty that we can identify and summarize them in Table~\ref{Tab:errs}.

\begin{table}
\caption{Uncertainty estimates.}
\label{Tab:errs}%
\begin{tabular}{c|c|c|c|c|c|c|c|c|c||c|c}
\hline \hline
Method & Quantity & Central Value & $s_{0}/\omega_{c}$ & $m_{b}/\Delta$ & NLO & $\mu$ & OPE & $\langle\bar{q}q\rangle$ & \multicolumn{2}{c|}{$1/m_{b}$} & Total err\tabularnewline
\hline 
\multirow{2}{*}{Full-QCDSR} & $\lambda_{+}/{\rm GeV}^{3}$ & $0.0148$ & $14\%$ & $11\%$ & $\approx33\%$ & $16\%$ & $\lesssim1\%$ & $2\%$ & \multicolumn{2}{c|}{-- --} & $\approx41\%$\tabularnewline
\cline{2-12}
 & $L_{1}$ & $-0.119$ & $34\%$ & $8\%$ & $\approx15\%$ & $12\%$ & $\lesssim1\%$ & $15\%$ & \multicolumn{2}{c|}{-- --} & $\approx43\%$\tabularnewline
\hline 
\multirow{2}{*}{HQET-SR} & $\lambda_{+}/{\rm GeV}^{3}$ & $0.0204$ & $8\%$ & $5\%$ & $\approx33\%$ & $10\%$ & $\lesssim1\%$ & $2\%$ & \multicolumn{2}{c|}{$12\%$} & $\approx38\%$\tabularnewline
\cline{2-12}
 & $L_{1}$ & $-0.159$ & $19\%$ & $12\%$ & $\approx15\%$ & $15\%$ & $\lesssim1\%$ & $11\%$ & \multicolumn{2}{c|}{$11\%$} & $\approx35\%$\tabularnewline
\hline \hline
\end{tabular}
\end{table}

\begin{figure}[htbp]
\centering
\includegraphics[width=0.6\linewidth]{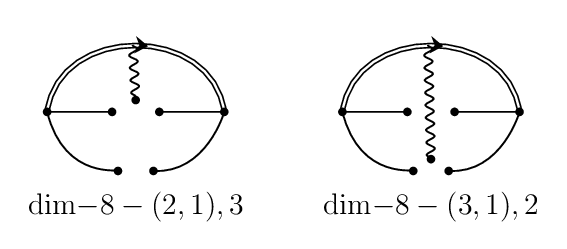}
\caption{The Feynman diagrams corresponding to the dimension-8 condensate contributions in the two-point correlation function.}
\label{fig:dim8-2pt}
\end{figure}

\begin{figure}[htbp]
\centering
\includegraphics[width=1\linewidth]{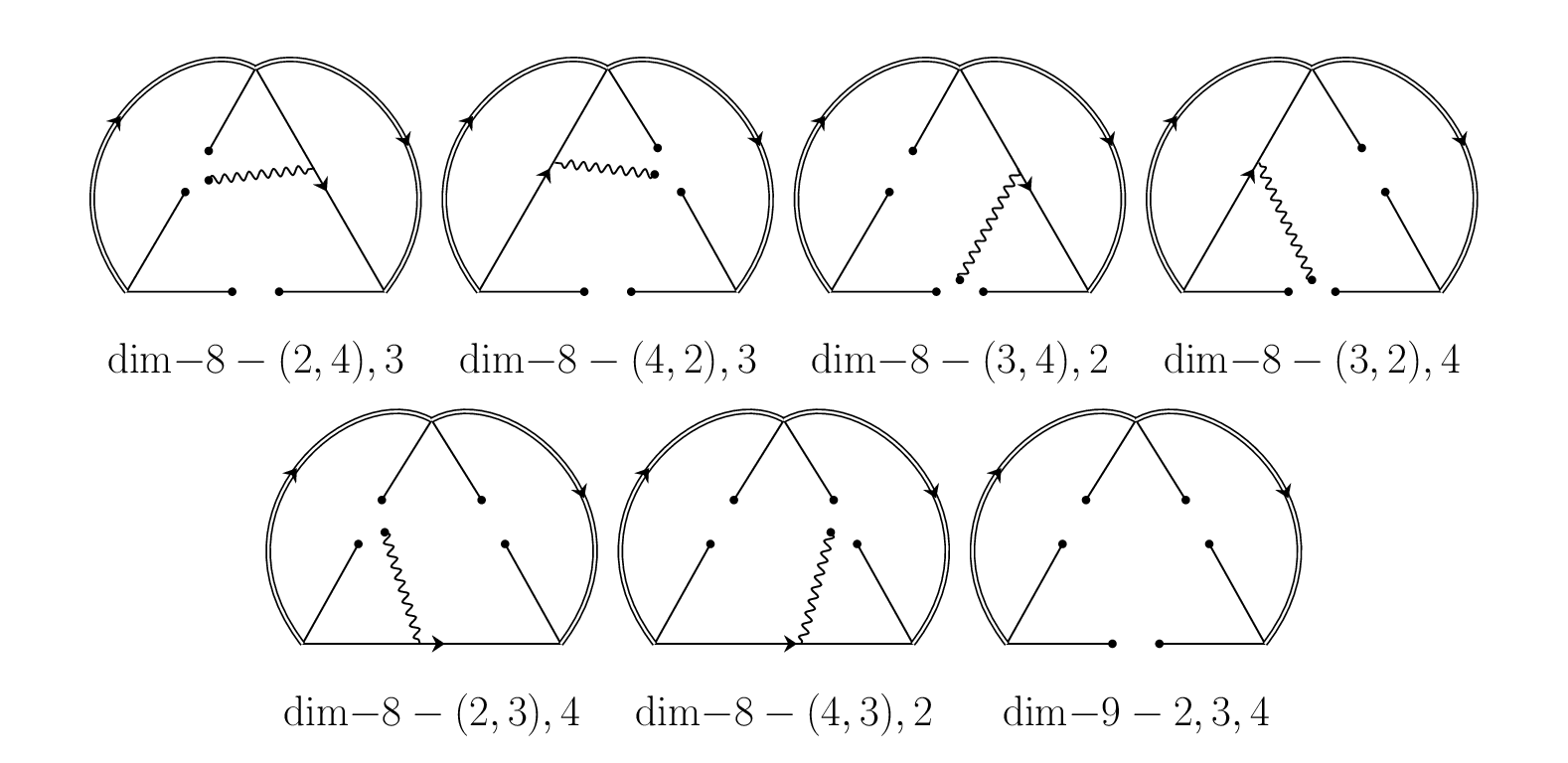}
\caption{The Feynman diagrams corresponding to the dimension-8 and dimension-9 condensate contributions in the three-point correlation function.}
\label{fig:dim89-3pt}
\end{figure}

\begin{enumerate}
\item Uncertainty from threshold parameters. 
See Figs.~\ref{fig:pole_residue} and \ref{fig:Li}.
\item $m_{b}/\Delta$ uncertainty. 
In the Full-QCDSR calculation, we include the uncertainty induced by the heavy-quark mass choice $m_{b}^{\rm pole}=4.80\pm0.06\,{\rm GeV}$. In the HQET-SR calculation, we include the uncertainty associated with the binding energy $\Delta=0.9\pm0.1\,{\rm GeV}$.
\item The next-to-leading order (NLO) uncertainty estimate. 
The NLO contribution can only be obtained through a full computation of the corresponding one-loop radiative corrections. The general strategy would be: perform IBP reduction of the loop diagrams, reduce them to master integrals, and finally evaluate the master integrals using cutting rules. We leave such a detailed analysis to future work. At present, we can only provide a very rough estimate of the NLO contribution. Following the method used in Ref.~\cite{Zhao:2021sje}, we estimate the full NLO correction using the $\alpha_{s}$ correction to the heavy-quark pole mass:
\begin{equation}
m_{b}^{\rm pole}=m_{b}(\mu)\left(1+\frac{4\alpha_{s}(\mu)}{3\pi}\right).
\end{equation}
By computing the relevant quantities at LO and NLO and comparing the results, we obtain an estimated uncertainty of about $33\%$ for the pole residue and about $15\%$ for $L_{1}$.  
For HQET-SR, a reliable NLO estimate is far beyond the scope of our present work; we can only very roughly assume that similar uncertainties may arise as in the Full-QCDSR case.
\item Scale-dependence uncertainty.  
In the full-QCDSR calculation, the central values are obtained in the pole-mass scheme, where the heavy-quark mass is scale independent. The scale dependence is estimated by evolving the condensates from $\mu=1\,{\rm GeV}$ to $\mu=0.5\,{\rm GeV}$, resulting in uncertainties of about $16\%$ for the pole residue and $12\%$ for $L_{1}$. The same procedure in HQET-SR leads to uncertainties of about $10\%$ and $15\%$, respectively.
\item OPE truncation uncertainty.  
To estimate this uncertainty, we further include the dimension-8 contribution for the pole residue, as shown in Fig.~\ref{fig:dim8-2pt}, and find that it vanishes due to special gamma-matrix structures. For $L_{1}$, we include both dimension-8 and dimension-9 contributions, as shown in Fig.~\ref{fig:dim89-3pt}, and find that the dimension-9 contribution also vanishes for the same reason, while the dimension-8 contribution is several orders of magnitude smaller than the perturbative and four-quark condensate contributions. 
In Ref.~\cite{Zhao:2021lzd}, we computed the dimension-4 gluon condensate contribution and found it to be several orders of magnitude smaller than the perturbative contribution. Based on this, we expect the dimension-7 contribution to also be much smaller than the perturbative contribution.
\item Uncertainty from condensate parameters. 
This uncertainty is straightforward to estimate. In this work, we have only used the quark condensate value $\langle\bar{q}q\rangle=(-0.24\pm0.01\,{\rm GeV})^{3}$.
\item $1/m_{b}$ power-correction uncertainty in HQET-SR.  
We uniformly set the continuum threshold parameter to $s_{0}=37\,{\rm GeV}^{2}$ for all full-QCDSR analyses and $\omega_{c}=1.28\,{\rm GeV}$ for all HQET-SR analyses. Taking typical values of the Borel parameters, we recalculate $\lambda_{+}$ and $L_{1}$. The results are summarized in Table~\ref{Tab:high_power_err}, from which, we conclude that the higher-order $1/m_b$ corrections missing in the HQET-SR contribute uncertainties of approximately $12\%$ and $11\%$ to $\lambda_{+}$ and $L_{1}$, respectively.
Note that here we estimate all higher-order power corrections beyond the leading $1/m_{b}$ term.
\end{enumerate}

\begin{table}[htbp]
\caption{To estimate the higher-order $1/m_{b}$ corrections in HQET-SR. 
We uniformly set the continuum threshold parameter to $s_{0}=37\,{\rm GeV}^{2}$ for all full-QCDSR analyses and $\omega_{c}=1.28\,{\rm GeV}$ for all HQET-SR analyses. Typical values of the Borel parameters are taken.}
\label{Tab:high_power_err} %
\begin{tabular}{c|c|c}
\hline \hline
Method & $\lambda_{+}/{\rm GeV}^{3}$ & $L_{1}$\tabularnewline
\hline 
Full-QCDSR & $0.0175$ & $-0.0845$\tabularnewline
\hline 
HQET-SR & $0.0198$ & $-0.0954$\tabularnewline
\hline \hline
\end{tabular}
\end{table}

\section{Conclusions}

Heavy quark expansion theory provides a good explanation for the lifetimes of weakly decaying heavy-flavor hadrons. The matrix elements of four-quark operators are the primary source for understanding the lifetime differences among hadrons containing the same heavy quark; therefore, precise calculations of these matrix elements are of significant theoretical importance.

In this work, we first obtain the leading-order results for the four-quark operator matrix elements within the framework of full QCD sum rules. At the OPE level, we consider the perturbative contribution and the four-quark condensate contribution. Our main emphasis is to highlight that the HQET sum rule results can be obtained by taking the heavy quark limit of the full QCD sum rules. For different dimension contributions, the calculations of taking the heavy quark limit are quite straightforward; however, for some contributions, they are rather nontrivial.

We also perform a detailed numerical analysis and uncertainty analysis in full-QCDSR and HQET-SR. The results show that the outcomes obtained within the two theoretical frameworks are consistent with each other. We also analyze in detail the sources of the large discrepancies between full-QCDSR and HQET-SR results reported in the literature.

This work contributes to a deeper understanding of the relationship between the two sum-rule approaches and is expected to facilitate future applications of QCD and HQET sum rules in the study of heavy-flavor hadrons. It is anticipated that the results of this work will provide insights for both practitioners of QCD sum rules and researchers in effective field theories.

\section*{Acknowledgements}

The authors are grateful to Prof.~Pietro Colangelo for valuable discussions. This work is supported in part by National Natural Science Foundation of China under Grants No.~12465018, 12065020. 

\appendix

\section{Spectral densities of the two-point correlation function in full-QCDSR}

\label{sec:rho-2pt}

In this subsection, we adopt the notation of Fig.~\ref{fig:feynman-2pt}, where $k_{1,2,3}$ and $m_{1,2,3}$ denote the momenta and masses of the $b$ quark and the two light quarks, respectively. The momentum $p$ is the total momentum introduced in the definition of the two-point correlation function in Eq.~(\ref{eq:2pt}), and $s\equiv p^{2}$. The Källén function is defined as
\begin{equation}
\lambda(a,b,c)\equiv a^{2}+b^{2}+c^{2}-2(ab+bc+ca).
\end{equation}

The calculation of the spectral densities of the perturbative contribution is the foundation of the two-point correlation function analysis. Here, we provide some details of the derivation. After applying the cutting rules, we obtain the following three-body phase space:
\begin{equation}
\Phi_{[1,2,3]}\equiv
\int d^{4}k_{1}d^{4}k_{2}d^{4}k_{3}
\delta(k_{1}^{2}-m_{1}^{2})
\delta(k_{2}^{2}-m_{2}^{2})
\delta(k_{3}^{2}-m_{3}^{2})
\delta^{4}(p-k_{1}-k_{2}-k_{3}).
\end{equation}
By inserting the identities
\begin{equation}
\int d^{4}k_{23}\delta^{4}(k_{23}-k_{2}-k_{3})=1,
\end{equation}
and
\begin{equation}
\int dm_{23}^{2}\delta(k_{23}^{2}-m_{23}^{2})=1,
\end{equation}
the three-body phase space can be rewritten as
\begin{equation}
\Phi_{[1,2,3]}
=\int dm_{23}^{2}\,\Phi_{[1,23]}\times\Phi_{[2,3]},
\end{equation}
where
\begin{align}
\Phi_{[1,23]} & \equiv
\int d^{4}k_{1}d^{4}k_{23}
\delta(k_{1}^{2}-m_{1}^{2})
\delta(k_{23}^{2}-m_{23}^{2})
\delta^{4}(p-k_{1}-k_{23}),
\nonumber\\
\Phi_{[2,3]} & \equiv
\int d^{4}k_{2}d^{4}k_{3}
\delta(k_{2}^{2}-m_{2}^{2})
\delta(k_{3}^{2}-m_{3}^{2})
\delta^{4}(k_{23}-k_{2}-k_{3}).
\end{align}
Therefore, the original three-body phase space is factorized into two two-body phase spaces. 

The final results for the spectral densities are given below.
The spectral densities of the perturbative contribution are
\begin{align}
\{\rho_{1}^{{\rm pert}},\rho_{2}^{{\rm pert}}\}
=&
\frac{3}{32\pi^{6}}
\int_{0}^{(\sqrt{s}-m_{1})^{2}}dm_{23}^{2}
\nonumber\\
&\times
\left\{
\frac{\pi^{2}m_{23}^{2}(s+m_{1}^{2}-m_{23}^{2})
\sqrt{\lambda(s,m_{1}^{2},m_{23}^{2})}}
{4s^{2}},
\frac{\pi^{2}m_{1}m_{23}^{2}
\sqrt{\lambda(s,m_{1}^{2},m_{23}^{2})}}
{2s}
\right\},
\end{align}
where
\begin{equation}
m_{23}^{2}\equiv(k_{2}+k_{3})^{2}.
\end{equation}

The spectral densities of the four-quark condensate contributions are
\begin{align}
\{\rho_{1}^{{\rm dim6}},\rho_{2}^{{\rm dim6}}\}
=
\frac{\langle\bar{q}q\rangle^{2}}{24}
\times\delta(s-m_{1}^{2})
\times\{4,4m_{1}\}.
\end{align}

\section{Spectral densities of the three-point correlation function in full-QCDSR}

\label{sec:rho-3pt}

In this subsection, we first specify the conventions for the quark momenta and masses. As shown in the perturbative diagram in Fig.~\ref{fig:feynman-3pt}, $k_{1,2,3,4,5}$ denote the momenta of the five quark propagators, respectively. Among them, $k_{1}$ and $k_{5}$ correspond to the momenta of the $b$ quarks, while the others represent the momenta of the light quarks. To keep the notation consistent, their corresponding masses are denoted by $m_{1,2,3,4,5}$.

The momenta $p_{1}$ and $p_{2}$ are introduced in the definition of the three-point correlation function in Eq.~(\ref{eq:correlator_3pt}), and we define
\begin{equation}
q=p_{1}-p_{2}.
\end{equation}
Furthermore, we denote
\begin{equation}
s_{1,2}\equiv p_{1,2}^{2}.
\end{equation}
Since we are interested in the forward-scattering matrix elements, the limit
\begin{equation}
q^{2}\to 0^{-}
\end{equation}
is understood throughout this section.

We provide some details on the calculation of the spectral densities of the perturbative contribution. After applying the cutting rules, we obtain the following generalized phase-space integral:
\begin{align}
{\cal P}_{[1,2,3,4,5]}\equiv &
\int d^{4}k_{1}d^{4}k_{2}d^{4}k_{3}d^{4}k_{4}d^{4}k_{5}
\delta(k_{1}^{2}-m_{1}^{2})
\delta(k_{2}^{2}-m_{2}^{2})
\delta(k_{3}^{2}-m_{3}^{2})
\delta(k_{4}^{2}-m_{4}^{2})
\delta(k_{5}^{2}-m_{5}^{2})
\nonumber\\
&\times
\delta^{4}(p_{1}-k_{1}-k_{2}-k_{3})
\delta^{4}(p_{2}-k_{3}-k_{4}-k_{5}).
\end{align}
By inserting the identities
\begin{equation}
\int d^{4}k_{12}\delta^{4}(k_{12}-k_{1}-k_{2})=1,
\end{equation}
\begin{equation}
\int dm_{12}^{2}\delta(k_{12}^{2}-m_{12}^{2})=1,
\end{equation}
and
\begin{equation}
\int d^{4}k_{45}\delta^{4}(k_{45}-k_{4}-k_{5})=1,
\end{equation}
\begin{equation}
\int dm_{45}^{2}\delta(k_{45}^{2}-m_{45}^{2})=1,
\end{equation}
the generalized phase-space integral can be factorized as
\begin{equation}
{\cal P}_{[1,2,3,4,5]}
=
\int dm_{12}^{2}dm_{45}^{2}\,
\Delta_{[12,3,45]}
\times
\Phi_{[1,2]}
\times
\Phi_{[4,5]},
\end{equation}
where
\begin{align}
\Delta_{[12,3,45]}\equiv &
\int d^{4}k_{12}d^{4}k_{3}d^{4}k_{45}
\delta(k_{12}^{2}-m_{12}^{2})
\delta(k_{3}^{2}-m_{3}^{2})
\delta(k_{45}^{2}-m_{45}^{2})
\nonumber\\
&\times
\delta^{4}(p_{1}-k_{12}-k_{3})
\delta^{4}(p_{2}-k_{3}-k_{45}),
\nonumber\\
\Phi_{[1,2]}\equiv &
\int d^{4}k_{1}d^{4}k_{2}
\delta(k_{1}^{2}-m_{1}^{2})
\delta(k_{2}^{2}-m_{2}^{2})
\delta^{4}(k_{12}-k_{1}-k_{2}),
\nonumber\\
\Phi_{[4,5]}\equiv &
\int d^{4}k_{4}d^{4}k_{5}
\delta(k_{4}^{2}-m_{4}^{2})
\delta(k_{5}^{2}-m_{5}^{2})
\delta^{4}(k_{45}-k_{4}-k_{5}).
\end{align}
Therefore, the generalized phase-space integral
${\cal P}_{[1,2,3,4,5]}$
is transformed into a convolution of two two-body phase spaces and one triangle phase space. 

The final results are given below. The spectral densities of the perturbative contribution are
\begin{align}
\{\rho_{A_{1}}^{{\rm pert}},\rho_{A_{2}}^{{\rm pert}},
\rho_{A_{3}}^{{\rm pert}},\rho_{A_{4}}^{{\rm pert}}\}
=&
\left(-\frac{3}{256\pi^{9}}\right)
\int_{m_{1}^{2}}^{s_{1}}dm_{12}^{2}
\int_{m_{1}^{2}}^{s_{2}}dm_{45}^{2}
\,
\Delta_{[12,3,45]}
\Phi_{[1,2]}
\Phi_{[4,5]}
\nonumber\\
&\times
\Bigg\{
4m_{1}
[m_{45}^{2}a_{122}B_{45}t_{2}
+m_{12}^{2}a_{454}B_{12}t_{1}],
\nonumber\\
&\quad
4[
m_{1}^{2}s_{1}a_{122}a_{454}t_{1}
+m_{12}^{2}m_{45}^{2}B_{12}B_{45}t_{2}],
\nonumber\\
&\quad
4[
m_{1}^{2}s_{2}a_{122}a_{454}t_{2}
+m_{12}^{2}m_{45}^{2}B_{12}B_{45}t_{1}],
\nonumber\\
&\quad
4m_{1}
[
m_{45}^{2}s_{1}a_{122}B_{45}t_{1}
+m_{12}^{2}s_{2}a_{454}B_{12}t_{2}]
\Bigg\},
\label{eq:rho_Ai_pert}
\end{align}
where
\begin{align}
a_{122}&=\frac{m_{12}^{2}-m_{1}^{2}+m_{2}^{2}}{2m_{12}^{2}},
\qquad
a_{454}=\frac{m_{45}^{2}+m_{4}^{2}-m_{5}^{2}}{2m_{45}^{2}},
\nonumber\\
B_{12}&=4b_{1200}+b_{1211},
\qquad
B_{45}=4b_{4500}+b_{4511},
\nonumber\\
b_{1200}&=
\frac{(m_{12}^{2})^{2}+(m_{1}^{2}-m_{2}^{2})^{2}
-2m_{12}^{2}(m_{1}^{2}+m_{2}^{2})}
{12(m_{12}^{2})^{2}},
\nonumber\\
b_{1211}&=
\frac{(m_{12}^{2})^{2}
-2(m_{1}^{2}-m_{2}^{2})^{2}
+m_{12}^{2}(m_{1}^{2}+m_{2}^{2})}
{6(m_{12}^{2})^{2}},
\nonumber\\
b_{4500}&=
\frac{(m_{45}^{2})^{2}+(m_{4}^{2}-m_{5}^{2})^{2}
-2m_{45}^{2}(m_{4}^{2}+m_{5}^{2})}
{12(m_{45}^{2})^{2}},
\nonumber\\
b_{4511}&=
\frac{(m_{45}^{2})^{2}
-2(m_{4}^{2}-m_{5}^{2})^{2}
+m_{45}^{2}(m_{4}^{2}+m_{5}^{2})}
{6(m_{45}^{2})^{2}},
\nonumber\\
t_{1}&=
-\frac{
\frac{1}{2}(-m_{12}^{2}+m_{3}^{2}+s_{1})s_{2}
-\frac{1}{4}(m_{3}^{2}-m_{45}^{2}+s_{2})(-q^{2}+s_{1}+s_{2})
}
{
-s_{1}s_{2}
+\frac{1}{4}(-q^{2}+s_{1}+s_{2})^{2}
},
\nonumber\\
t_{2}&=
-\frac{
\frac{1}{2}s_{1}(m_{3}^{2}-m_{45}^{2}+s_{2})
-\frac{1}{4}(-m_{12}^{2}+m_{3}^{2}+s_{1})(-q^{2}+s_{1}+s_{2})
}
{
-s_{1}s_{2}
+\frac{1}{4}(-q^{2}+s_{1}+s_{2})^{2}
}.
\end{align}
The results for the two-body phase spaces and the triangle phase space are
\begin{align}
\Phi_{[1,2]}
&=
\frac{\pi\sqrt{\lambda(m_{12}^{2},m_{1}^{2},m_{2}^{2})}}
{2m_{12}^{2}},
\nonumber\\
\Phi_{[4,5]}
&=
\frac{\pi\sqrt{\lambda(m_{45}^{2},m_{4}^{2},m_{5}^{2})}}
{2m_{45}^{2}},
\nonumber\\
\Delta_{[12,3,45]}
&=
\frac{\pi}
{2\sqrt{\lambda(2s_{1}+2s_{2}-q^{2},s_{1},s_{2})}}
\nonumber\\
&\times
\Theta\Big[
-s_{1}s_{2}q^{2}
-(m_{45}^{2})^{2}s_{1}
-(m_{12}^{2})^{2}s_{2}
-(m_{3}^{2})^{2}q^{2}
\nonumber\\
&\qquad
+(-s_{1}+s_{2}+q^{2})
(m_{45}^{2}s_{1}+m_{12}^{2}m_{3}^{2})
\nonumber\\
&\qquad
+(s_{1}-s_{2}+q^{2})
(m_{12}^{2}s_{2}+m_{3}^{2}m_{45}^{2})
\nonumber\\
&\qquad
+(s_{1}+s_{2}-q^{2})
(m_{3}^{2}q^{2}+m_{12}^{2}m_{45}^{2})
\Big].
\end{align}
The calculations of the two-body phase spaces are relatively straightforward and can usually be found in standard quantum field theory textbooks. For the triangle phase space, it is convenient to perform the calculation in the rest frame of $p_{1}+p_{2}$. Although Eq.~(\ref{eq:rho_Ai_pert}) can be further simplified, we do not perform this simplification in order to keep the expression more transparent.

The spectral densities of the four-quark condensate contribution from dim-6-2,3 are
\begin{align}
\{
\rho_{A_{1}}^{{\rm dim623}},
\rho_{A_{2}}^{{\rm dim623}},
\rho_{A_{3}}^{{\rm dim623}},
\rho_{A_{4}}^{{\rm dim623}}
\}
=&
\left(
-\frac{\langle\bar{q}q\rangle^{2}}
{192\pi^{3}}
\right)
\times
\delta(s_{1}-m_{1}^{2})
\nonumber\\
&\times
\pi m_{1}
\left(1-\frac{m_{1}^{2}}{s_{2}}\right)^{2}
\times
\left\{
1,m_{1},
\frac{s_{2}}{m_{1}},
s_{2}
\right\}.
\end{align}
The spectral densities of the four-quark condensate contribution from dim-6-4,3 can be obtained in an analogous way.

The spectral densities of the four-quark condensate contribution from dim-6-2,4 are
\begin{align}
\{
\rho_{A_{1}}^{{\rm dim624}},
\rho_{A_{2}}^{{\rm dim624}},
\rho_{A_{3}}^{{\rm dim624}},
\rho_{A_{4}}^{{\rm dim624}}
\}
=&
\left(
-\frac{\langle\bar{q}q\rangle^{2}}
{192\pi^{3}}
\right)
\times
\Theta
\left[
-q^{2}(m_{1}^{2}-s_{1})(m_{1}^{2}-s_{2})
-m_{1}^{2}(s_{1}-s_{2})^{2}
\right]
\nonumber\\
&\times
\frac{2\pi}
{\lambda^{3/2}(s_{1},s_{2},q^{2})}
\nonumber\\
&\times
\Big\{
m_{1}
[
2m_{1}^{2}q^{2}
-q^{2}(s_{1}+s_{2})
+(s_{1}-s_{2})^{2}
],
\nonumber\\
&\qquad
(m_{1}^{4}-s_{1}s_{2})
(q^{2}+s_{1}-s_{2}),
\nonumber\\
&\qquad
(m_{1}^{4}-s_{1}s_{2})
(q^{2}-s_{1}+s_{2}),
\nonumber\\
&\quad
-m_{1}
\{
m_{1}^{2}
[
(s_{1}-s_{2})^{2}
-q^{2}(s_{1}+s_{2})
]
+2q^{2}s_{1}s_{2}
\}
\Big\}.
\end{align}
It should be emphasized that the $q^2$ dependence in the theta 
function should never be omitted. The underlying reason will become clear from 
the detailed derivation in the next section.

\section{Heavy-quark limit of the spectral density of dim-6-2,4}

\label{sec:HQL-624}

Up to an overall coefficient
$(-\frac{\langle\bar{q}q\rangle^{2}}{192\pi^{3}})$,
the spectral density of the dim-6-2,4 contribution in full-QCDSR is given by
\begin{equation}
\rho(s_{1},s_{2},q^{2})
=
\frac{
8\pi m_{1}q^{2}(m_{1}^{4}-s_{1}s_{2})
}
{
\left[
q^{4}+(s_{1}-s_{2})^{2}
-2q^{2}(s_{1}+s_{2})
\right]^{3/2}
}
\Theta
\left[
-q^{2}(m_{1}^{2}-s_{1})(m_{1}^{2}-s_{2})
-m_{1}^{2}(s_{1}-s_{2})^{2}
\right],
\end{equation}
where $\rho = \{M_{-}^{2},M_{-},M_{-},1\}.\{\rho^{A_{1}},\rho^{A_{2}},\rho^{A_{3}},\rho^{A_{4}}\}$, and we have adopted the approximation $M_{-}\approx m_{1}$. 

To take the heavy-quark limit, we introduce the variables
\begin{equation}
s_{i}=(m_{1}+\sigma_{i})^{2},
\qquad
\Sigma=\frac{\sigma_{1}+\sigma_{2}}{2},
\qquad
\Delta=\sigma_{1}-\sigma_{2},
\end{equation}
and rewrite
\begin{equation}
q^{2}=-Q^{2},
\qquad Q>0 .
\end{equation}

In the heavy-quark limit, one obtains
\begin{equation}
m_{1}^{4}-s_{1}s_{2}
=
-4m_{1}^{3}\Sigma+\mathcal{O}(m_{1}^{2}),
\end{equation}
and
\begin{equation}
q^{4}+(s_{1}-s_{2})^{2}
-2q^{2}(s_{1}+s_{2})
=
4m_{1}^{2}(\Delta^{2}+Q^{2})
+\mathcal{O}(m_{1}).
\end{equation}
However, for the argument of the $\Theta$ function, instead of taking the heavy-quark limit, we simply rewrite it as
\begin{equation}
-q^{2}(m_{1}^{2}-s_{1})(m_{1}^{2}-s_{2})-m_{1}^{2}(s_{1}-s_{2})^{2}
=m_{1}^{2}(2m_{1}+\sigma_{1}+\sigma_{2})^{2}(a^{2}-\Delta^{2}),
\end{equation}
where
\begin{equation}
a=\frac{Q}{m_{1}}\sqrt{\sigma_{1}\sigma_{2}}
\frac{\sqrt{(2m_{1}+\sigma_{1})(2m_{1}+\sigma_{2})}}
{2m_{1}+\sigma_{1}+\sigma_{2}}.
\end{equation}

For the coefficient of the $\Theta$ function, we retain only the leading term in the $1/m_{1}$ expansion. The spectral density then reduces to
\begin{equation}
\rho(s_{1},s_{2},q^{2})
=
4\pi m_{1}\Sigma
\frac{Q^{2}}
{(\Delta^{2}+Q^{2})^{3/2}}
\Theta(a^{2}-\Delta^{2}).
\end{equation}
Since
\begin{equation}
\frac{a}{Q}\to0
\end{equation}
in the heavy-quark limit, within the support region
$|\Delta| < a$ one can make the replacement
\begin{equation}
(\Delta^{2}+Q^{2})^{3/2}
\rightarrow Q^{3}.
\end{equation}
Therefore, the spectral density further simplifies to
\begin{equation}
\rho(s_{1},s_{2},q^{2})
=
\frac{4\pi m_{1}\Sigma}{Q}
\Theta(a^{2}-\Delta^{2}).
\end{equation}
Since $|\Delta| < a$ and $a\to 0$, we have $\Delta \to 0$. 
Using the rectangular representation of the delta function:
\begin{equation}
\lim_{a\rightarrow0}
\frac{1}{2a}
\Theta(a^{2}-\Delta^{2})
=
\delta(\Delta),
\end{equation}
we finally obtain 
\begin{equation}
\rho(s_{1},s_{2},q^{2})
\to
\frac{4\pi m_{1}\Sigma}{Q}
2a\delta(\Delta)
\to 
8\pi\sigma_{1}\sigma_{2}
\delta(\sigma_{1}-\sigma_{2}).
\end{equation}



\begin{thebibliography}{1}

\bibitem{LHCb:2018nfa}
R.~Aaij \textit{et al.} [LHCb],
Phys. Rev. Lett. \textbf{121}, no.9, 092003 (2018)
doi:10.1103/PhysRevLett.121.092003
[arXiv:1807.02024 [hep-ex]].

\bibitem{Tanabashi:2018oca}
M.~Tanabashi \textit{et al.} [Particle Data Group],
Phys. Rev. D \textbf{98}, no.3, 030001 (2018)
doi:10.1103/PhysRevD.98.030001

\bibitem{LHCb:2021vll}
R.~Aaij \textit{et al.} [LHCb],
Sci. Bull. \textbf{67}, no.5, 479-487 (2022)
doi:10.1016/j.scib.2021.11.022
[arXiv:2109.01334 [hep-ex]].

\bibitem{Belle-II:2022plj}
F.~J.~Abudinen \textit{et al.} [Belle-II],
Phys. Rev. D \textbf{107}, no.3, L031103 (2023)
doi:10.1103/PhysRevD.107.L031103
[arXiv:2208.08573 [hep-ex]].

\bibitem{Cheng:2018rkz}
H.~Y.~Cheng,
JHEP \textbf{11}, 014 (2018)
doi:10.1007/JHEP11(2018)014
[arXiv:1807.00916 [hep-ph]].

\bibitem{Gratrex:2022xpm}
J.~Gratrex, B.~Meli{\'c} and I.~Ni{\v{s}}and{\v{z}}i{\'c},
JHEP \textbf{07}, 058 (2022)
doi:10.1007/JHEP07(2022)058
[arXiv:2204.11935 [hep-ph]].

\bibitem{Cheng:2023jpz}
H.~Y.~Cheng and C.~W.~Liu,
JHEP \textbf{07}, 114 (2023)
doi:10.1007/JHEP07(2023)114
[arXiv:2305.00665 [hep-ph]].

\bibitem{Zhao:2021lzd}
Z.~X.~Zhao, X.~Y.~Sun, F.~W.~Zhang and Z.~P.~Xing,
Eur. Phys. J. C \textbf{84}, no.1, 48 (2024)
doi:10.1140/epjc/s10052-023-12299-9
[arXiv:2101.11874 [hep-ph]].

\bibitem{Colangelo:1996ta}
P.~Colangelo and F.~De Fazio,
Phys. Lett. B \textbf{387}, 371-378 (1996)
doi:10.1016/0370-2693(96)01049-0
[arXiv:hep-ph/9604425 [hep-ph]].

\bibitem{Shi:2019hbf}
Y.~J.~Shi, W.~Wang and Z.~X.~Zhao,
Eur. Phys. J. C \textbf{80}, no.6, 568 (2020)
doi:10.1140/epjc/s10052-020-8096-2
[arXiv:1902.01092 [hep-ph]].

\bibitem{Zhao:2020mod}
Z.~X.~Zhao, R.~H.~Li, Y.~L.~Shen, Y.~J.~Shi and Y.~S.~Yang,
Eur. Phys. J. C \textbf{80}, no.12, 1181 (2020)
doi:10.1140/epjc/s10052-020-08767-1
[arXiv:2010.07150 [hep-ph]].

\bibitem{Zhao:2021sje}
Z.~X.~Zhao, X.~Y.~Sun, F.~W.~Zhang, Y.~P.~Xing and Y.~T.~Yang,
Phys. Rev. D \textbf{108}, no.11, 116008 (2023)
doi:10.1103/PhysRevD.108.116008
[arXiv:2103.09436 [hep-ph]].

\bibitem{Xing:2021enr}
Z.~P.~Xing and Z.~X.~Zhao,
Eur. Phys. J. C \textbf{81}, no.12, 1111 (2021)
doi:10.1140/epjc/s10052-021-09902-2
[arXiv:2109.00216 [hep-ph]].

\bibitem{Zhang:2023nxl}
S.~Q.~Zhang and C.~F.~Qiao,
Phys. Rev. D \textbf{108}, no.7, 074017 (2023)
doi:10.1103/PhysRevD.108.074017
[arXiv:2307.05019 [hep-ph]].

\bibitem{Zhang:2024ick}
S.~Q.~Zhang, X.~H.~Zhang and C.~F.~Qiao,
JHEP \textbf{06}, 122 (2024)
doi:10.1007/JHEP06(2024)122
[arXiv:2402.15088 [hep-ph]].

\bibitem{Zhang:2024asb}
S.~Q.~Zhang and C.~F.~Qiao,
Phys. Rev. D \textbf{110}, no.11, 114040 (2024)
doi:10.1103/PhysRevD.110.114040
[arXiv:2411.15857 [hep-ph]].

\bibitem{Lu:2025gol}
J.~Lu, G.~L.~Yu, D.~Y.~Chen, Z.~G.~Wang and B.~Wu,
Eur. Phys. J. C \textbf{85}, no.12, 1382 (2025)
doi:10.1140/epjc/s10052-025-15110-z
[arXiv:2508.11900 [hep-ph]].

\bibitem{Yu:2026tbk}
G.~L.~Yu, Z.~G.~Wang, J.~Lu, B.~Wu, P.~Yang and Z.~Zhou,
Phys. Rev. D \textbf{113}, no.7, 076013 (2026)
doi:10.1103/ghm9-2lmd
[arXiv:2601.06427 [hep-ph]].

\bibitem{Black:2024bus}
M.~Black, M.~Lang, A.~Lenz and Z.~W{\"u}thrich,
JHEP \textbf{04}, 081 (2025)
doi:10.1007/JHEP04(2025)081
[arXiv:2412.13270 [hep-ph]].

\bibitem{Shuryak:1981fza}
E.~V.~Shuryak,
Nucl. Phys. B \textbf{198}, 83-101 (1982)
doi:10.1016/0550-3213(82)90546-6

\bibitem{Colangelo:1995qp}
P.~Colangelo, C.~A.~Dominguez, G.~Nardulli and N.~Paver,
Phys. Rev. D \textbf{54}, 4622-4628 (1996)
doi:10.1103/PhysRevD.54.4622
[arXiv:hep-ph/9512334 [hep-ph]].

\bibitem{Li:2020ejs}
H.~n.~Li and H.~Umeeda,
Phys. Rev. D \textbf{102}, 114014 (2020)
doi:10.1103/PhysRevD.102.114014
[arXiv:2006.16593 [hep-ph]].

\bibitem{Xiong:2022uwj}
A.~S.~Xiong, F.~S.~Yu, Y.~Zheng and T.~Wei,
[arXiv:2211.13753 [hep-th]].

\bibitem{Zhao:2024drr}
Z.~X.~Zhao, Y.~P.~Xing and R.~H.~Li,
Eur. Phys. J. C \textbf{84}, no.10, 1105 (2024)
doi:10.1140/epjc/s10052-024-13452-8
[arXiv:2407.09819 [hep-ph]].

\bibitem{Khoze:1983yp}
V.~A.~Khoze and M.~A.~Shifman,
Sov. Phys. Usp. \textbf{26}, 387 (1983)
doi:10.1070/PU1983v026n05ABEH004398

\bibitem{Bigi:1991ir}
I.~I.~Y.~Bigi and N.~G.~Uraltsev,
Phys. Lett. B \textbf{280}, 271-280 (1992)
doi:10.1016/0370-2693(92)90066-D

\bibitem{Bigi:1992su}
I.~I.~Y.~Bigi, N.~G.~Uraltsev and A.~I.~Vainshtein,
Phys. Lett. B \textbf{293}, 430-436 (1992)
[erratum: Phys. Lett. B \textbf{297}, 477-477 (1992)]
doi:10.1016/0370-2693(92)90908-M
[arXiv:hep-ph/9207214 [hep-ph]].

\bibitem{Blok:1992hw}
B.~Blok and M.~A.~Shifman,
Nucl. Phys. B \textbf{399}, 441-458 (1993)
doi:10.1016/0550-3213(93)90504-I
[arXiv:hep-ph/9207236 [hep-ph]].

\bibitem{Blok:1992he}
B.~Blok and M.~A.~Shifman,
Nucl. Phys. B \textbf{399}, 459-476 (1993)
doi:10.1016/0550-3213(93)90505-J
[arXiv:hep-ph/9209289 [hep-ph]].

\bibitem{Neubert:1997gu}
M.~Neubert,
Adv. Ser. Direct. High Energy Phys. \textbf{15}, 239-293 (1998)
doi:10.1142/9789812812667{\_}0003
[arXiv:hep-ph/9702375 [hep-ph]].

\bibitem{Uraltsev:1998bk}
N.~Uraltsev,
Proc. Int. Sch. Phys. Fermi \textbf{137}, 329-409 (1998)
doi:10.3254/978-1-61499-222-6-329
[arXiv:hep-ph/9804275 [hep-ph]].

\bibitem{Bigi:1995jr}
I.~I.~Y.~Bigi,
[arXiv:hep-ph/9508408 [hep-ph]].

\bibitem{Lenz:2014jha}
A.~Lenz,
Int. J. Mod. Phys. A \textbf{30}, no.10, 1543005 (2015)
doi:10.1142/S0217751X15430058
[arXiv:1405.3601 [hep-ph]].

\bibitem{Wang:2010fq}
Z.~G.~Wang,
Eur. Phys. J. C \textbf{68}, 479-486 (2010)
doi:10.1140/epjc/s10052-010-1365-8
[arXiv:1001.1652 [hep-ph]].

\bibitem{Penin:1998kx}
A.~A.~Penin and A.~A.~Pivovarov,
Nucl. Phys. B \textbf{549}, 217-241 (1999)
doi:10.1016/S0550-3213(99)00182-0
[arXiv:hep-ph/9807421 [hep-ph]].

\bibitem{Colangelo:2000dp}
P.~Colangelo and A.~Khodjamirian,
doi:10.1142/9789812810458{\_}0033
[arXiv:hep-ph/0010175 [hep-ph]].

\bibitem{ParticleDataGroup:2024cfk}
S.~Navas \textit{et al.} [Particle Data Group],
Phys. Rev. D \textbf{110}, no.3, 030001 (2024)
doi:10.1103/PhysRevD.110.030001

\end{thebibliography}
\end{document}